\documentclass[final,3p,times]{elsarticle}

\usepackage{hyperref}
\usepackage[utf8]{inputenc}
\usepackage{array}
\usepackage{amssymb}
\usepackage{amsmath}
\usepackage{color}
\newcommand{\spsmsA}{Univ. Grenoble Alpes, INAC-SPSMS,
                     F-38000 Grenoble, France}
\newcommand{\spsmsB}{CEA, INAC-SPSMS, F-38000 Grenoble, France}

\def\be{\begin{equation}}
\def\ee{\end{equation}}

\journal{Physica E}
\begin{document}

\begin{frontmatter}

\title{A computational approach to quantum noise in time-dependent nanoelectronic devices}

\author{Benoit Gaury}
\author{Xavier Waintal}
\address{\spsmsA,}
\address{\spsmsB}

\begin{abstract}
We derive simple expressions that relate the noise and correlation properties of a general time-dependent quantum conductor
to the wave functions of the system. The formalism provides a practical route for numerical calculations of quantum noise in an externally driven system. We illustrate the approach with
numerical calculations of the noise properties associated to a voltage pulse applied on a one-dimensional conductor. The methodology is hower fully general and can be used for a large class of mesoscopic conductors.
\end{abstract}

\end{frontmatter}

\section{Introduction} Among the many contributions of Markus B\"uttiker to the
field of mesoscopic physics (now best known as nanoelectronics), his pioneer work
on time-dependent phenomena was particularly dear to him. He was insistent in
pointing out the role of displacement currents (required to restore current
conservation), the need for a theory that preserved ``gauge invariance" and more generally built up the general framework and concepts to address this physics.
While the key theoretical works were performed in the early 90s~\cite{Buttiker_LC,Buttiker_dynamic_conductance,Buttiker_capacitors,Buttiker_admittances,Buttiker_TDcurrent_partition},
the corresponding experiments were difficult (GHz physics at mK temperatures) so that more than 10 years elapsed before the first
quantum RC circuit could actually be measured~\cite{RC_Kirchhoff}. The field has considerably matured since then, with the latest
generation of experiments performed directly in the time domain~\cite{dubois_minimal-excitation_2013}.

 Markus visited Saclay on March 2006 
which is when the senior author of this paper was introduced to the subject. At that time, although the analytical theory was well developed, its computational counterpart was still in its infancy (see~\cite{gaury_numerical_2014} for a short history). We now have
very effective numerical tools~\cite{gaury_numerical_2014}, with a computational effort linear both in time and system size, so that numerical complexity is no longer an issue~\cite{gaury_dynamical_2014} to simulate time-dependent systems. 
In this article, we would like to extend these tools to
calculate another quantity, also very dear to Markus, namely the quantum fluctuations of observables. Quantum noise is a quantity which
is not only sensitive to the wave aspect of quantum transport but also to
particle statistics~\cite{blanter2000}. We shall see that its numerical calculation requires some care in order to disentangle
the relevant contributions from the large background already present at equilibrium.

\section{The Pauli principle in driven mesoscopic systems}
Let us consider a general mesoscopic system described by a time dependent (externally driven) quadratic Hamiltonian,
\begin{equation}
    \mathrm{\hat{\textbf{H}}}(t) =
    \sum_{i,j} \mathrm{\textbf{H}}_{ij}(t) c^{\dagger}_{i}c_{j}
\end{equation}
where $c^{\dagger}_{i}$ ($c_{j}$) are the usual Fermionic creation (annihilation)
operators of a one-particle state on site $i$. The site index $i$ typically labels space as well
as spin, orbital (s,p,d,f) and/or Nambu (for superconductors) degrees of freedom. The system is open, 
i.e. consists of a finite time-dependent central part connected to infinite (stationary) electrodes. We further suppose
that the Hamiltonian is time independent for $t\le 0$ and we switch on the time-dependent part (voltage pulses, light\dots ) at
$t>0$.

For $t\le 0$, the solution to this problem is well known: one introduces the stationary scattering states $\Psi_{\alpha E}^{st}$ (labeled by their energy $E$ and lead mode $\alpha$) which diagonalize the one-body Hamiltonian,
\begin{equation}
 \sum_{j} \mathrm{\textbf{H}}_{ij}(0) \Psi_{\alpha E}^{st}(j) = E \Psi_{\alpha E}^{st}(i)
\end{equation}
and build many-body Slater determinants from these one-body states. For quantities such as the current, this amounts to filling up the
energy $E$ with probability $f_\alpha(E)$ (Fermi function of the lead to which
channel $\alpha$ belongs) which leads to the celebrated Landauer formula for the
conductance. Within our notations, the current between sites $i$ and $j$ reads,
 \begin{equation}
I_{ij} =-\frac{2e}{h} \  {\rm Im } \sum_\alpha  \ \int dE\ f_\alpha (E)
[\Psi^{st}_{\alpha E}(i)]^* \mathrm{\textbf{H}}_{ij}(0)\Psi^{st}_{\alpha E}(j).
\end{equation}
This approach is by now so standard that $\Psi_{\alpha E}^{st}$ can be obtained
directly from open source softwares such as
the Kwant package~\cite{kwant2014}. The generalization of the above approach to time-dependent phenomena is in fact straightforward:
one simply follows the evolution of the scattering states upon switching on the time-dependent perturbation and solves,
\begin{equation}
\label{eq:sch}
i\hbar \frac{\partial}{\partial t} \Psi_{\alpha E}(i,t) = \sum_{j} \mathrm{\textbf{H}}_{ij}(t) \Psi_{\alpha E}(j,t) 
\end{equation}
with the initial condition
\begin{equation}
\label{eq:ini}
  \Psi_{\alpha E}(i, t=0) =  \Psi_{\alpha E}^{st}(i).
\end{equation}
The time-dependent current is now given by,
\begin{equation}
\label{eq:cur}
I_{ij}(t) =-\frac{2e}{h} \  {\rm Im } \sum_\alpha  \ \int dE\ f_\alpha (E)
[\Psi_{\alpha E}(i,t)]^* \mathrm{\textbf{H}}_{ij}(t)\Psi_{\alpha E}(j,t).
\end{equation}
Equation (\ref{eq:sch}) is transparent physically but not directly useful for numerical computations as the wave-function
$\Psi_{\alpha E}(i,t)$ is a vector of infinite size. However, by studying the {\it deviation} between $\Psi_{\alpha E}(i,t)$
and $\Psi_{\alpha E}^{st}(i)e^{-i E t}$ one obtains a finite vector amenable to a numerical solution~\cite{gaury_numerical_2014}.
In order to prove that Equations (\ref{eq:sch}), (\ref{eq:ini}) and
(\ref{eq:cur}) lead to the correct generalization of the Landauer formula,
Ref.~\cite{gaury_numerical_2014} proved its mathematical equivalence with the
Wingreen-Meir~\cite{Wingreen-Meir_1992} 
approach based on the Keldysh formalism. One can also prove that the $\Psi_{\alpha E}(i,t)$ are well defined scattering states, i.e.
have the correct structure of superposition of incoming and outgoing states in the leads. The Pauli principle is fully enforced within this scheme, as the initial orthogonality relations of the states are preserved by the unitary evolution of Schr\"odinger equation.
It is interesting to contrast the above set of equations with studies
where a (usually Gaussian) wavepacket is propagated through the system. {\it In fine}, the dynamical equation solved is actually the same.
There are however two crucial differences: the initial boundary condition (delocalized overt the infinite system in our case) and the final integration over energy which restores the fermionic statistics. The approach presented here is fully many-body and treats Pauli principle exactly. 

\section{Quantum Noise}

We now discuss how the approach outlined above can
be generalized to calculate the noise properties of a general mesoscopic system
subject to time-dependent perturbation. We will focus in particular on the variance of the total
number of particles $\hat n_\mu$ sent through an electrode $\mu$. This quantity is of special interest to us
as it is relatively easy to measure experimentally (upon sending repeated pulses it is essentially a d.c. measurement
as opposed to much more challenging high frequency measurements) and it is {\it
conserved and gauge invariant} in Markus' sense~\cite{gaury_numerical_2014}.

\subsection{General expressions}
The current operator is defined as a
sum of the local currents flowing through a cross section corresponding to an electrode $\mu$
(in practice $\mu$ is the collection of hopping elements that connect the central system to the electrode)

\be
\hat{I}_{\mu}(t) = \sum_{\langle i,j \rangle \in \mu} 
\mathrm{\textbf{H}}_{ij}(t) c_i^{\dagger}(t)c_j(t) - \mathrm{\textbf{H}}_{ji}(t)
c_j^{\dagger}(t)c_i(t),
\ee
and we define the current-current correlation function as
\be
    S_{\mu \nu}(t,t') = \left(\hat{I}_{\mu}(t) - \langle \hat{I}_{\mu}(t) \rangle \right)
              \times \left(\hat{I}_{\nu}(t') - \langle \hat{I}_{\nu}(t') \rangle \right).
\label{defcor}
\ee
The calculation of Eq.~(\ref{defcor}) requires the statistical average of
products of four fermionic operators which is directly obtained using the Wick theorem and the expressions for the
one-particle Green's function that were derived in~\cite{gaury_numerical_2014},
\begin{align}
\label{G<}
&\langle c^{\dagger}_j(t') c_i(t) \rangle = \sum_{\alpha} \int \frac{dE}{2\pi}
f_{\alpha}(E) \Psi_{\alpha E}^*(j, t') \Psi_{\alpha E}(i,t)  \\
 \label{G>}
 &\langle c_i(t) c^{\dagger}_j(t') \rangle = \sum_{\alpha} \int \frac{dE}{2\pi}
 (1 - f_{\alpha}(E)) \Psi_{\alpha E}(i,t) \Psi_{\alpha E}^*(j, t'). \\
\end{align}
One obtains,
\begin{align}
    S_{\mu \nu}(t,t') = 
    \sum_{\alpha, \beta}& \int \frac{dE}{2\pi} \int \frac{dE'}{2\pi}
    f_{\alpha}(E)(1-f_{\beta}(E'))  I_{\mu,EE'}(t) \left[I_{\nu,EE'}(t')\right]^*,
\label{correlator}
\end{align}
with the quantity $I_{\mu,EE'}(t)$ closely related to the initial current operator
\begin{align}
\label{EEcurrent}
    I_{\mu,EE'}(t) = \sum_{\langle i,j \rangle \in \mu}\Big[
    \Psi_{\beta E'}^*(i,t) H_{ij}(t) \Psi_{\alpha E}(j, t)
    -\Psi_{\beta E'}^*(j,t) H_{ji}(t)\Psi_{\alpha E}(i, t) \Big].
\end{align}
Equations (\ref{correlator}) relates the typical output of a time-dependent simulation (right-hand side) to the correlation properties
(left-hand side). To proceed, we focus on the total number of transmitted
particles $\hat n_\mu$ over a duration $\Delta$, 
\be
    \hat{n}_\mu = \int_{-\Delta/2}^{\Delta/2} dt\ \hat{I}_{\mu}(t).
\ee
With these notations, the average and variance of $\hat{n}_\mu$ are given by,
\be
\langle \hat n_\mu\rangle =  \sum_{\alpha} \int \frac{dE}{2\pi}\  
    f_{\alpha}(E) N_{EE}
\ee 
\be
\label{eq:bare}
     \mathrm{var}(\hat{n}_\mu)= 
    \sum_{\alpha, \beta} \int \frac{dE}{2\pi} \int \frac{dE'}{2\pi}\ 
    f_{\alpha}(E)(1-f_{\beta}(E')) |N_{EE'}|^2,
\ee
with 
\be
N_{EE'} = \int_{-\Delta/2}^{\Delta/2} dt\ I_{\mu,EE'}(t). 
\ee
For a finite $\Delta$, the above equations are well defined. However, as we shall see, the noise is totally dominated by the equilibrium noise which diverges at large $\Delta$ so that the numerical computation of the excess noise (the quantity which is usually measured) 
can become problematic. Ultimately, we would like to consider the limit where $\Delta$ is infinite. 
To proceed, we  separate the equilibrium physics from the time-dependent one and introduce $\bar\Psi_{\alpha E}(i, t)$ which
measures how the wave-function deviates from its stationary solution: 
\be
\Psi_{\alpha E}(i, t)    =   \Psi_{\alpha E}^{st}(i, t)e^{-iEt}  + \bar\Psi_{\alpha E}(i, t)
\ee
$\bar\Psi_{\alpha E}(i, t)$ is actually the direct output of the techniques discussed in~\cite{gaury_numerical_2014}. With
these notations, one can perform the integration over time (in the limit of large $\Delta$) and write the noise in terms of well
behaved (converging) integrals. The variance of $\hat n_\mu$ now reads
\be
\label{var3}
    \mathrm{var}(\hat{n}_{\mu}) =  \sigma^2_{st} \ \Delta  +  2\sigma_{mix} + \bar \sigma^2  + O(1/\Delta)
\ee
where the three numbers $\sigma^2_{st}$, $\sigma_{mix}$ and $\bar \sigma^2$ are defined as,
\begin{align}
    \sigma^2_{st} &= \sum_{\alpha, \beta} \int \frac{dE}{2\pi}\ f_{\alpha}(E)(1-f_{\beta}(E))
    |I_{\mu,EE}(0)|^2 \\
    \sigma_{mix} &= \sum_{\alpha, \beta} \int \frac{dE}{2\pi}\ f_{\alpha}(E)(1-f_{\beta}(E))
     \mathrm{Re}[\bar{N}_{EE}^* I_{\mu,EE}(0)] \\
    \bar \sigma^2 &= \sum_{\alpha, \beta} \int \frac{dE}{2\pi}\ \frac{dE'}{2\pi}
    f_{\alpha}(E)[1-f_{\beta}(E')] |\bar{N}_{EE'}|^2
\end{align}
in term of 
\begin{align}
    \bar{N}_{EE'} = \int_{-\infty}^{\infty} dt\ \left[I_{\mu,EE'}(t) -
    I_{\mu,EE'}(0)e^{-i(E-E')t}\right].
\end{align}
The convergence of the time integrals is assured if the time-dependent perturbation is of finite duration.
The first term of Eq.~(\ref{var3}) is simply the stationary noise of the problem (including the Johnson-Nyquist noise 
and some shot noise contributions) while the two last terms form the extra noise due to the time-dependent perturbation.
Eq.~(\ref{var3}) is very convenient for practical calculations as the diverging stationary noise can be directly subtracted from
the calculation and one does not suffer from numerical inaccuracies. Eq.~(\ref{var3}) is the main result of this article.
\begin{figure}[h]
\centering
\includegraphics[width=0.4\textwidth]{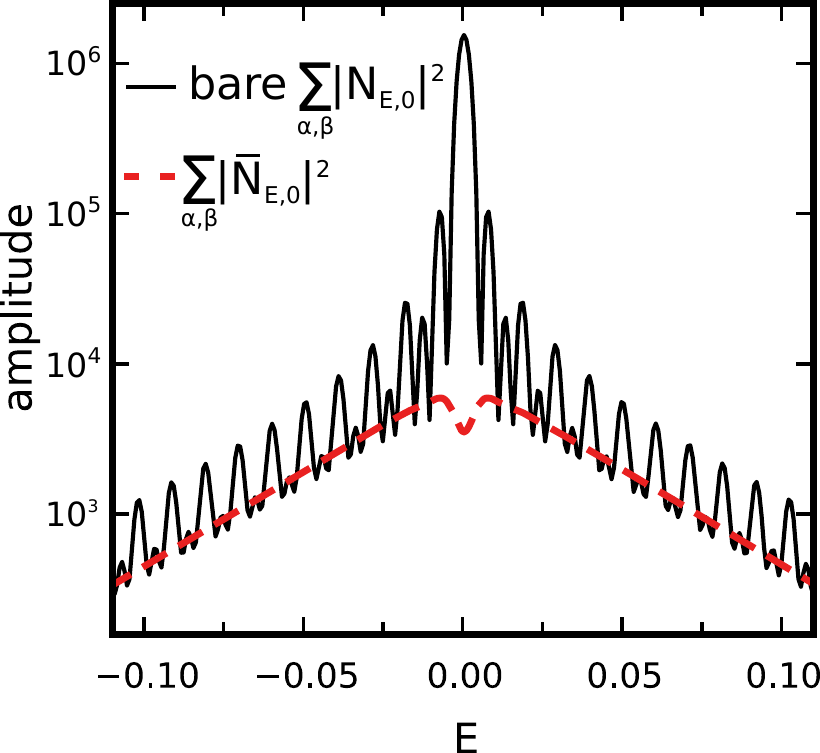}
\caption{Comparison of  the amplitudes of the bare $\sum_{\alpha, \beta} |N_{E,0}|^2$
(black full line) with  $\sum_{\alpha, \beta} |\bar N_{E,0}|^2$ where the
diverging term has been subtracted (red dash line) as a function of energy.}
\label{techfig}
\end{figure}
The two quantities that need to be integrated, ${N}_{EE'}$ (bare) and
$\bar{N}_{EE'}$ (after subtraction of the stationary state), are plotted for
illustration in Fig.~\ref{techfig} for a voltage pulse sent through a simple one-dimensional infinite wire with a barrier in the middle (see~\cite{gaury_dynamical_2014} for the precise microscopic model). 
One observes that the interesting contribution of $\bar{N}_{EE'}$ is completely
overshadowed by the stationary contribution (that contains a term of the form
$\sin (\Delta E)/E$ which eventually converges to a dirac function) so that
practical calculations are greatly facilitated by the use of Eq.~(\ref{var3})
instead of Eq.~(\ref{eq:bare}).

\subsection{Application to a one-dimensional conductor}
\begin{figure}
\centering
  \includegraphics[width=0.7\textwidth]{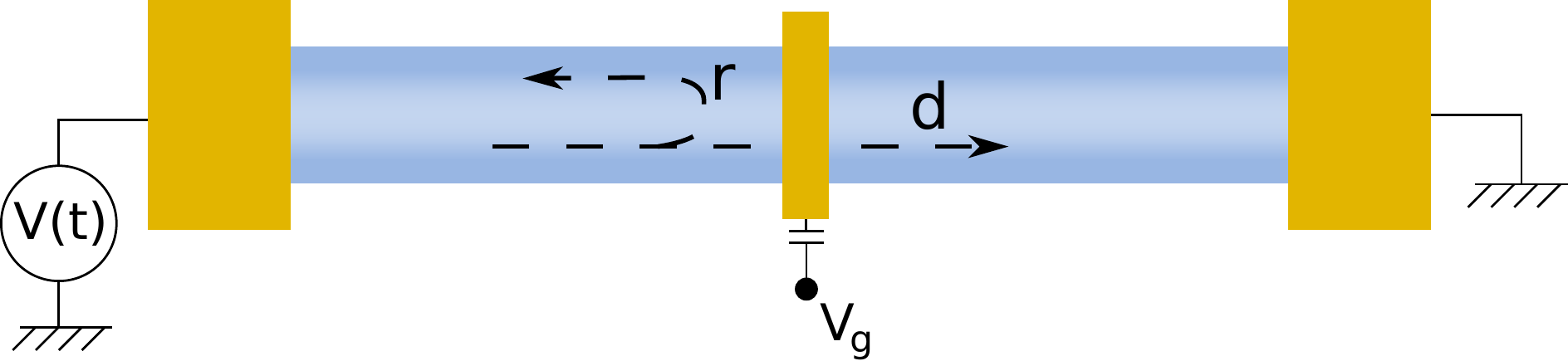} 
    \caption{Schematic of our one-dimensional wire connected to two electrodes.
    The gate voltage $V_g$ creates a barrier with transmission (reflection)
    amplitude $d$ ($r$). A voltage pulse is applied on the left electrode while the right one is grounded.
    }
    \label{wire}
\end{figure}

The study of the noise associated to a voltage pulse in a quantum conductor
was pioneered some time ago by the work of Levitov and coworkers~\cite{Lorentzian_pulses,Levitov1996}
but the first experiments were only performed very recently~\cite{Glattli}. Let us quickly revisit this
issue. We consider a one-dimensional wire  with
a central barrier of transmission (reflection) amplitude $d$ ($r$), and we
apply a voltage pulse $V(t)$ on the left electrode while the right one is
grounded, see Fig.~\ref{wire} for a schematic. 
Assuming a linear spectrum $E = vk$ and neglecting the energy dependence
of the barrier, the wave-function on the right of the barrier for
electrons coming from the left and right electrodes takes the form 
\begin{align}
    &\Psi_{L,E}(n,t) = \frac{d}{\sqrt{v}} e^{ikn -iEt -i\phi(t)} \\
    &\Psi_{R,E}(n,t) = \frac{e^{-ikn-iEt}}{\sqrt{v}} +
    \frac{r}{\sqrt{v}}e^{ikn-iEt}
\end{align}
where the two indices $L,R$ label the electrode (Left or Right), $v=\partial E / \partial k$ is the group velocity 
and $\phi(t)=\int^t du\ eV(u)/\hbar$. Inserting these wave-functions into Eq.~(\ref{correlator}) and defining $\epsilon=E-E'$,
we obtain
\begin{align}
    &S_{RR}(t,t') = \int \frac{dE}{2\pi} \int \frac{d\epsilon}{2\pi}\ 
                  f(E)(1-f(E-\epsilon)) 
                \left(2D^2 + 2D(1-D)\cos[\phi(t)-\phi(t')]\right) e^{i\epsilon(t-t')}
\end{align}
where $D=|d|^2$ is the transmission probability of the tunneling barrier. Performing the successive
integrals over the times $t$ and $t'$, and over the energy $E$ one obtains for the variance, 
\begin{align}
    \mathrm{var}(\hat{n}_\mu) &= \frac{k_BT_eD^2}{\pi}\Delta 
  + D(1-D) \int \frac{d\epsilon}{4\pi^2} 
 |K(\epsilon)|^2 \epsilon \coth\Big(\frac{\epsilon}{2k_BT_e}\Big)
\label{varn}
\end{align}
where $K(\epsilon)=\int dt\ e^{i\phi(t) +i\epsilon t}$ is the amplitude of
probability that the voltage pulse leads to a change $\epsilon$ of the energy, 
$k_B$ is the Boltzmann constant and $T_e$ the temperature.
The second term of Eq.~(\ref{varn}) is equal to Eq.~(17) of
Ref~\cite{Lorentzian_pulses} up to the barrier factor $D(1-D)$. 
The partitioning of Eq.~(\ref{varn}) is however not very transparent as the (diverging) equilibrium noise is actually contained
in both terms of the equation. In fact, in the absence of pulse and for infinite $\Delta$,
$K(\epsilon) = \delta(\epsilon)$ so that the second term of Eq.~(\ref{varn}) contains the square of a dirac function
and is ill defined.
Let us  now compute the three contributions $\sigma^2_{st}$, $\sigma_{mix}$ and $\bar \sigma^2$
separately. The deviation from equilibrium of the wave-function takes the form
\begin{align}
    &\bar \Psi_{L,E}(n,t) = \frac{d}{\sqrt{v}}(e^{ -i\phi(t)}-1) e^{ikn -iEt} \\
    &\bar \Psi_{R,E}(n,t) = 0.
\end{align}
and one obtain after some algebra,\begin{align}
    \sigma^2_{st} &= \frac{k_BT_eD}{\pi} \\
    \sigma_{mix} &=  \frac{k_BT_eD(1-D)}{\pi} \Bigg[\int_{-\Delta/2}^{\Delta/2}
    dt\ \cos(\phi(t)) - \Delta \Bigg] \\
    \bar \sigma^2 + 2\sigma_{mix} &= D(1-D)\int \frac{d\epsilon}{4\pi^2}\ 
    |K(\epsilon)|^2 \epsilon \coth\Big(\frac{\epsilon}{2k_BT_e}\Big) 
     - \frac{k_BT_eD(1-D)}{\pi} \Delta
\end{align}
One can easily check that the summation of all these terms gives back the result
of Eq.~(\ref{varn}) as it should. The advantage of this form is that the different integrals
are now naturally regularized while the Johnson-Nyquist contribution is well separated from the rest. 
Equation (\ref{varn}) was further analyzed by Levitov {\it et al.} in Ref.~\cite{Levitov1996} and
one obtains a variance of the form (at zero temperature), 
\be
\label{an}
 \mathrm{var}(\hat{n}_{\mu}) =  D(1-D) \bar{n}
  + D(1-D) a \sin^2 (\pi\bar n)
\ee
where $a$ is a numerical factor and $\bar{n}=\int dt\ eV(t)/h$ is the number of electrons sent by the voltage pulse. The oscillating behavior in $\sin^2(\pi\bar{n})$
has been extensively discussed in the context of recent
experiments~\cite{Glattli}. Similar oscillations are also predicted for the current itself in the context of
interferometers~\cite{gaury_dynamical_2014}.
\begin{figure}[t]
    \centering
    \includegraphics[width=0.4\textwidth]{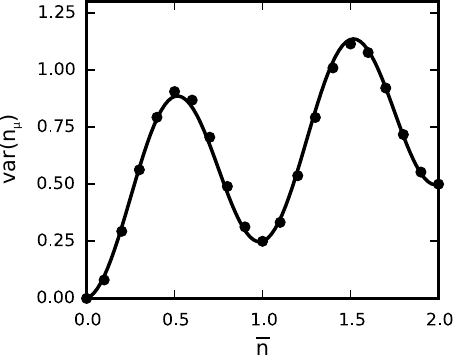} 
    \caption{Voltage pulse across a simple barrier. Variance of the number of transmitted particles at zero temperature as a function of the number of injected particles. The symbols correspond to the numerical data obtained by direct integration of the wave-function. The full line corresponds to Eq.(\ref{an} with $a\approx 3$. The barrier transmission is set to $D=0.5$.
    }
    \label{noise-0K}
\end{figure}

\begin{figure}[h]
\centering
      \includegraphics[width=0.4\textwidth]{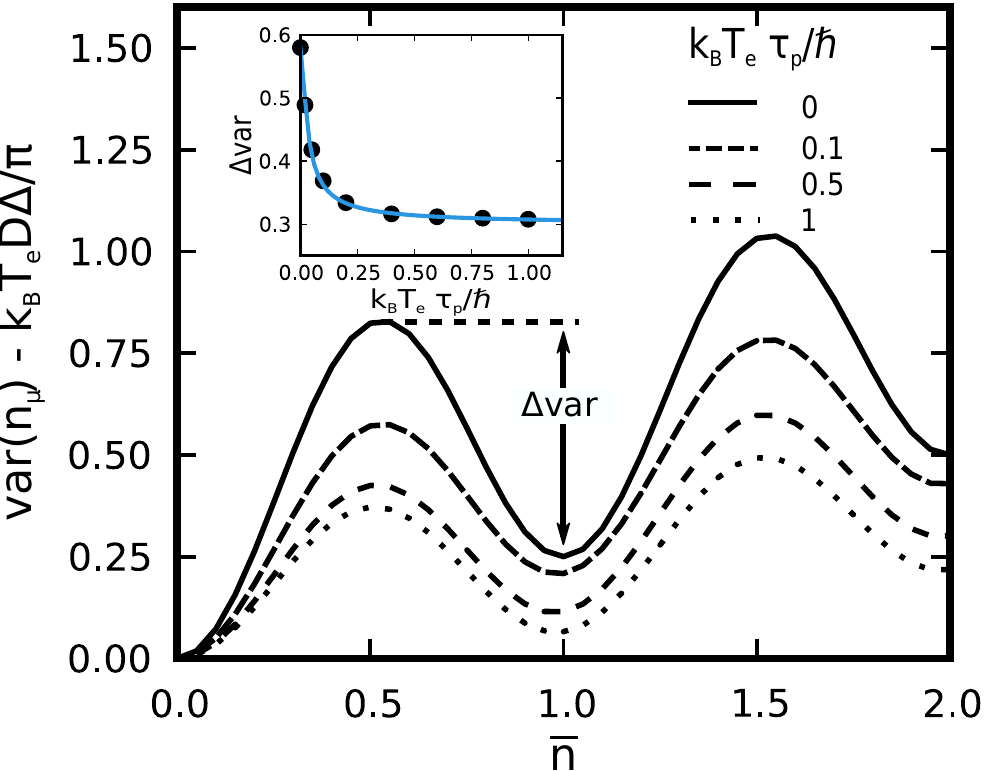} 
    \caption{Variance of the number of transmitted particles where the
    Johnson-Nyquist noise has been subtracted as a function of the number of
    injected particles. $\Delta \mathrm{var}$ is the difference of the noise for
    $\bar n =0.5$ and $\bar n =1$. Inset: $\Delta \mathrm{var}$ as a function of temperature
    in the unit of $\hbar/\tau_p$. The dots are numerical data, and the
    continuous line is a fit obtained with $\Delta \mathrm{var} = a \exp(-b \hbar /
    (\tau_p k_BT_e))$. The barrier transmission is set to $D=0.5$.
    }
    \label{noise-kTW}
\end{figure}

To illustrate our numerical procedure, we calculate the variance of the number of particles transmitted in the
case of Lorentzian voltage pulse, $V(t)=V_p/(1+(2t/\tau_p)^2)$, with $V_p$ the amplitude and $\tau_p$ the
duration of the pulse (and $\bar n\propto V_p/\tau_p$). 
Fig.~\ref{noise-0K} shows $var (\hat n_\mu)$ as a function of 
$\bar n$ at zero temperature and for a semi-transmitting barrier $D=0.5$. We get a perfect match with the analytical prediction.
Fig.~\ref{noise-kTW} shows the noise $var(\hat n_\mu)$ at various
temperatures $T_e$ after subtraction of the equilibrium noise. 
For $\bar n = 1$ the noise is minimum, i.e. equal to the
shot noise created by an equivalent d.c. current. This is a peculiar feature of the Lorentzian shape
which sends a unique electron (without creating additional electron-hole pairs
that would cause extra noise). The amplitude of the oscillations shows a rapid
decrease with temperature as shown in inset.

\section{Conclusion}

We have presented an extension of the numerical technique of~\cite{gaury_numerical_2014} that allows one to compute
the quantum noise properties of a conductor. The method naturally separates the stationary noise 
from the extra contributions coming from time-dependent perturbations such as voltage pulses. 
Once the wave-functions are known for different values of
the energy, the calculation of the noise simply amounts to a numerical integral
which can be performed as a post process. As an application we have considered
the simple case of a one-dimensional system but the numerical technique is in no
way limited to this case. In particular, the large class of systems which have
already been considered for the time-dependent simulations (which includes the
quantum Hall effect, electronic interferometers, conventional and topological
superconductors...) can now be studied for their noise properties.
Higher moments, correlations or fluctuations of other observables (such as charge fluctuations) can also be obtained following the same lines.  

{\it Acknowledgments.}
This work is funded by the ERC consolidator grant
MesoQMC. We thank C. Bauerle, C. Glattli, L. Glazmann, F. Portier and P. Roche for useful discussions.

\section*{References}

\bibliography{TKwantNoise}

\end{document}